\documentstyle[12pt]{article}
\newcommand{\newsection}{
\setcounter{equation}{0}
\section}
\def\appendix#1{
  \addtocounter{section}{1}
  \setcounter{equation}{0}
  \renewcommand{\thesection}{\Alph{section}}
  \section*{Appendix \thesection\protect\indent #1}
  \addcontentsline{toc}{section}{Appendix \thesection\ \ \ #1}
  }
\newcommand{\rf}[1]{(\ref{#1})}
\newcommand{\beq}{\begin{equation}}
\newcommand{\eeq}{\end{equation}}
\newcommand{\bea}{\begin{eqnarray}}
\newcommand{\eea}{\end{eqnarray}}

\renewcommand{\b}{\beta}
\renewcommand{\a}{\alpha}

\newcommand{\om}{\omega}

\newcommand{\oh}{\frac{1}{2}}

\newcommand{\dg}{\dagger}

\newcommand{\tr}{{\rm Tr}\;}

\newcommand{\cim}{\oint_{C}\frac{d\omega}{4\pi i}}
\newcommand{\ci}{\oint_{C}\frac{d\omega}{4\pi i}\;\frac{\omega
V'(\omega)}{p^2-\omega^2}}

\begin{document}
\topmargin 0pt
\oddsidemargin 5mm
\headheight 0pt
\headsep 0pt
\topskip 9mm

\addtolength{\baselineskip}{0.20\baselineskip}
\hfill    NBI-HE-92-45

\hfill June 1992
\begin{center}

\vspace{36pt}
{\large \bf
Higher Genus Correlators for the Complex Matrix Model
}

\vspace{36pt}

{\sl J. Ambj\o rn and C.F. Kristjansen}

\vspace{12pt}

 The Niels Bohr Institute\\
Blegdamsvej 17, DK-2100 Copenhagen \O , Denmark\\

\vspace{24pt}

and

\vspace{24pt}

{\sl Yu.\ M.\ Makeenko}

\vspace{12pt}

 Institute for Theoretical and Experimental Physics\\
117259 Moscow, Russia\\

\end{center}

\vspace{24pt}

\vfill

\begin{center}
{\bf Abstract}
\end{center}
We describe an iterative scheme which allows us to calculate any multi-loop
correlator for the complex matrix model to any genus using only the first in
the chain of loop equations.
The method works for a completely general potential and the
results contain no explicit reference to the couplings.
The genus $g$ contribution to the $m$--loop correlator
depends on a finite number of parameters, namely at most $4g-2+m$.
We find the generating functional explicitly up to genus three.
We show as well that the model is equivalent to an external field problem
for the complex matrix model with a logarithmic potential.

\vspace{12pt}

\noindent
\vspace{24pt}

\vfill

\newpage

\newsection{Introduction}
In the matrix model approach to 2D quantum gravity one
traditionally considers the hermitian matrix model. However, the complex
matrix model is just as well suited for this purpose. The complex matrix
model differs from the hermitian one in the discretized surface
language by giving rise to only "chequered surfaces"~\cite{Morris1}. Such
short distance properties of the surfaces should however not matter in
the continuum limit, and one do find that the two models lead to the
same string equation~\cite{Morris1}.

The complex matrix model is in several aspects
more appealing than the hermitian one. First it is possible to express
in one formula all multi--loop correlators for genus zero and it might
very well be possible to do so also for higher genera~\cite{jjm}.
Secondly it is possible to find a version of the complex matrix model
for which the string equation has a pole--free solution with the same
perturbative expansion as the the usual Painlev\'{e}
solution~\cite{Morris2}. There is
a possible drawback of the complex matrix model, though. Its Virasoro
generators seem to be ill defined in the continuum limit~\cite{metc}.
However one can identify in the correlators of the theory the term from
which these divergencies originate. This term poses no problems for the
double scaling limit of the correlators.

After having introduced the basic concepts in sections~2 and~3,
we put forward in section~4
a conjecture about the form of the genus $g$
contribution to the generating functional of the complex matrix model.
The conjecture is proven by induction. An outline of the proof
including the first step is given in section~5. The details can be
found in Appendix A. The proof
provides us with an iterative scheme by means of which it is possible to
calculate explicitly any multi--loop correlator to any genus. The method
works for a completely general potential and the results contain no
explicit reference to the couplings. The genus $g$ contribution to the
$m$--loop correlator depends on a finite number of parameters, namely at
most $4g-2+m$. Results for $g=2$ and $g=3$ are presented in section~6.
Everything is based only on the first in the chain of loop
equations.

Another iterative procedure based on work by Migdal~\cite{Migdal}
has been advocated by David~\cite{david}.
However, this procedure involves the entire chain
of loop equations and is applicable in practice only to potentials with
a finite (small) number of couplings. It was applied to the hermitian
matrix model with a quadratic and a cubic potential in~\cite{jm}.
In reference~\cite{jcm} an iterative procedure similar to the one described
in this paper was developed for the hermitian matrix model. Since it was
shown \cite{cm} that the usual hermitian matrix model is equivalent to the
so-called Kontsevich--Penner model \cite{cm1}, the iterative procedure can
be formulated in the language of the latter. We show in  Appendix~B that
an analogous property holds for the usual complex matrix model which is
actually  equivalent to an external field problem for the complex matrix model
with a logarithmic potential
(an analog of the Kontsevich--Penner model) but we formulate the iterative
procedure independently of this.

\newsection{Basic definitions}
The complex matrix model is defined by the partition function
\beq
Z=\int d\phi^{\dg} d\phi\exp(-NV(\phi^{\dg} \phi)) \label{partition}
\eeq
where the integration is over complex $N\times N$ matrices and
\beq
V(\phi^{\dg} \phi)=\sum_{j=1}^{\infty}\frac{g_j}{j}\tr(\phi^{\dg}\phi)^j\,.
\label{potential}
\eeq
We introduce the generating functional
\beq
W(p)=\frac{1}{N}\sum_{k=0}^{\infty}\langle \tr(\phi^{\dg}\phi)^{k}\rangle
/p^{2k+1}
\eeq
and the $n$--loop correlator $(n\geq2)$
\beq
W(p_1,\ldots,p_n)=N^{n-2}\sum_{k_1,\ldots,k_n=1}^{\infty}
\langle \tr(\phi^{\dg}\phi)^{k_1} \ldots
\tr(\phi^{\dg}\phi)^{k_n})\rangle_C
/p_1^{2k_1+1}\ldots p_n^{2k_n+1}\,.
\eeq
The multi--loop correlators can be obtained from the generating
functional by application of the loop insertion operator,
$\frac{d}{dV(p)}$,:
\beq
W(p_1,\ldots, p_n)=\frac{d}{dV(p_n)}\;\frac{d}{dV(p_{n-1})}\ldots
\frac{d}{dV(p_2)}\;W(p_1)
\eeq
where
\beq
\frac{d}{dV(p)}\equiv
-\sum_{j=1}^{\infty}\frac{j}{p^{2j+1}}\frac{d}{dg_{j}}\;.
\eeq

It is possible to show that the model defined by~\rf{partition}
and~\rf{potential} is equivalent to
an external field problem for the complex matrix model
with a logarithmic potential
(an analog of the hermitian Kontsevich--Penner model)
in much the same way as was the case for the hermitian matrix
model \cite{cm}. We defer the discussion of this to Appendix~B since the
iterative procedure can be formulated independently of this.

\newsection{The loop equation}
The first in the chain of loop equations can conveniently be written
as \cite{Mak0}
\beq
\ci W(\om) = (W(p))^2 +\frac{1}{N^2} W(p,p)
\label{loop}
\eeq
where $C$ is a curve which encloses singularities of $W(p)$ and
$V(\om)=\sum_{j}g_j\om^{2j}/j $.
With the normalization chosen above the genus expansion for the
correlators reads
\beq
W(p_1,\ldots,p_n)=\sum_{g=0}^{\infty}\frac{1}{N^{2g}}\;
W_{g}(p_1,\ldots,p_n) \hspace{1.0cm} (n\geq 1) \;.
\label{genus}
\eeq
To leading order in $1/N^2$ one can thus ignore the last term
in~\rf{loop} and one finds
\beq
W_{0}(p)=\oh \ci \left(\frac{p^2+c}{\om^2+c}\right)^{1/2}
\eeq
where c is given by
\beq
\cim\; \frac{\om V'(\om)}{(\om^2+c)^{1/2}}=2 \;.
\label{norm}
\eeq

Inserting the genus expansion~\rf{genus} in~\rf{loop} it is seen that
$W_g(p)\,,\;g\geq1$ obeys the following equation
\beq
\left\{\hat{K}-2W_{0}(p)\right\}W_g(p)=\sum_{g'=1}^{g-1}
W_{g'}(p)\;W_{g-g'}(p)
+\frac{d}{d V(p)}W_{g-1}(p) \label{loopg}
\eeq
where  $\hat{K}$ is a linear operator, namely
\beq
\hat{K}f(p)\equiv \ci f(\om)\;.
\eeq
In equation~\rf{loopg} $W_g(p)$ is expressed entirely in terms of
$W_{g_i}(p),\;\;g_i<g$. This indicates that one should be able to calculate
$W_{g}(p)$ for any finite genus $g$ starting from $W_{0}(p)$. In the next
section we describe an iterative procedure which makes this possible.

\newsection{The conjecture}
To characterize the matrix model potential we introduce, instead of the
couplings $g_j$, the moments $I_n$ and $M_k$ defined by
\bea
M_{k}&=&\cim \frac{V'(\om)}{w^{2k+1}(w^2+c)^{1/2}}, \hspace{1.0cm} k\geq
0\\
I_{n}&=&\cim \frac{\om V'(\om)}{(w^2+c)^{n+1/2}},   \hspace{1.7cm} n\geq 0\;.
\eea
The normalization condition~\rf{norm} is then simply, $I_{0}=2$, and the
$k^{th}$ multicritical point is reached when
\beq
I_1 = I_2 =\ldots =I_{k-1}=0,\hspace{1.0cm} I_k \neq 0\;.
\eeq

It is shown below that $W_g(p)$ can be expressed entirely in terms of
the moments  and that, for a given finite genus $g$, $W_g(p)$ depends
only on a finite number of these. On the contrary $W_g(p)$ depends
on all couplings $g_j$. Thus working with the moments instead of the
couplings might facilitate calculations considerably.

Furthermore we introduce the basis vectors $\chi^{(n)}(p)$ and
$\Psi^{(n)}(p)$ characterized by
\bea
\left\{\hat{K}-2W_0(p)\right\}\chi^{(n)}(p) &= & \frac{1}{(p^2+c)^n}
\label{Kchi}\\
\left\{\hat{K}-2W_0(p)\right\}\Psi^{(n)}(p) &=&\frac{1}{p^{2n}}\;.
\label{KPsi}
\eea
It can be shown that $\chi^{(n)}(p)$ and $\Psi^{(n)}(p)$ can be expressed as
\bea
\chi^{(n)}(p)&=&\frac{1}{I_1}\left\{\Phi^{(n)}(p)
         -\sum_{k=1}^{n-1}\chi^{(k)}(p)I_{n-k+1}
          \right\}
\label{chi}\\
\Psi^{(n)}(p)&=&\frac{1}{M_0}\left\{\Omega^{(n)}(p)
         -\sum_{k=1}^{n-1}\Psi^{(k)}(p)M_{n-k}
          \right\}
\label{Psi}
\eea
where
\bea
\Phi^{(n)}(p)&=&\frac{1}{(p^2+c)^{n+1/2}}\label{Phi}\\
\Omega^{(n)}(p)&=&\frac{1}{p^{2n}(p^2+c)^{1/2}}\label{Omega}\;.
\eea

We now put forward the following conjecture
\beq
W_g(p)=\sum_{n=1}^{3g-1}A_g^{(n)}\chi^{(n)}(p)
       +\sum_{m=1}^g D_g^{(m)} \Psi^{(m)}(p)\;.
\label{conjecture}
\eeq
The coefficient $A_g^{(n)}$ is a sum of terms of the form
\beq
a_{g}^{(n)} = I_{\a_1}I_{\a_2}\ldots I_{\a_k}
                M_{\b_1}M_{\b_2}\ldots M_{\b_l}\;f(c,M_0,I_1)
                \label{coeff}
\eeq
where
\beq
\a_1,\a_2,\ldots,\a_k \in [2,3g-n] \hspace{1.0cm}
\b_1,\b_2,\ldots,\b_l \in [1,g]\;.
\eeq
Let us for an expression of the type~\rf{coeff} denote by $H_I$ and
$H_M$
the degree of homogeneity in the $I$'s and $M$'s, respectively,
\beq
H_I\equiv \sum_{j=1}^{k}(\a_j-1), \hspace{1.0cm} H_M\equiv
\sum_{j=1}^{l}\b_j\;.
\label{homs}
\eeq
Then the following inequalities hold for the $a_g^{(n)}$'s
\beq
H_I\;(a_g^{(n)})\leq 3g-n-1, \hspace{1.0cm} H_M\;(a_g^{(n)})\leq g \;.
\label{hom1}
\eeq
The structure of the coefficients $D_g^{(m)}$ is similar to that of
the $A_g^{(n)}$'s. However, the  $f$'s will in
general be different and for the $d_g^{(m)}$'s we have
\beq
\a_1,\a_2,\ldots,\a_k \in [2,3g] \hspace{1.0cm}
\b_1,\b_2,\ldots,\b_l \in [1,g-m]
\eeq
and
\beq
H_I(d_g^{(m)})\leq 3g-1; \hspace{1.0cm} H_M(d_g^{(m)})\leq g-m\;.
\label{hom2}
\eeq

The homogeneity requirements become more transparent if one considers
the double scaling limit. For the $m^{th}$ multicritical model the double
scaling limit is obtained by fixing the ratio between any given coupling
and say $g_1$ to its critical value and setting
\bea
p^2&=&z_c+a\pi\\
c&=&-z_c+a\Lambda^{1/m}
\eea
where $\Lambda$ is the cosmological constant. The $I$'s then scale as
\beq
I_n\sim a^{m-n},\hspace{1.0cm} n\in[1,m-1]\,.
\eeq
Furthermore it is well known that the genus $g$ contribution to the
generating functional has the following scaling behaviour
\beq
W_g(\pi,\Lambda)\sim a^{(1-2g)(m+1/2)-1}
\eeq
with the exception of $W_0(\pi,\Lambda)$ which also contains a non
scaling part. From~\rf{chi}~--~\rf{Omega} it
follows that the basis vectors behave as
\beq
\chi^{(n)}\sim a^{-m-n+1/2}, \hspace{1.0cm} \Psi^{(n)} \sim a^{-1/2}\;.
\eeq

Bearing in mind that the coefficients $D_g^{(m)}$ and $A_g^{(n)}$ are of
the form~\rf{coeff} and noting that the $I_1$ dependence of
$f$ is also powerlike, including however negative powers, one
finds for the $D$'s and $A$'s
\bea
A_g^{(n)}:\hspace{0.5cm} && \sum_{j}(m-\a_j)\leq m(2-2g)+n-g-1 \\
D_g^{(n)}:\hspace{0.5cm} && \sum_{j}(m-\a_j)\leq (1-2g)(m+1/2)-1/2 \;.
\eea
Here we have used the same notation as in~\rf{homs}. However $\a_j=1$
is allowed and negative powers of $I_1$ give negative contributions to
the sums. Since the generating functional away from the double scaling
limit should look the same for all multicritical models the $D$'s and
the $A$'s must satisfy the following conditions
\bea
A_g^{(n)}:\hspace{0.5cm} & \sum_{j}1 \leq 2-2g \hspace{1.0cm}
&\sum_j(\a_j-1) \leq 3g-n-1
\label{homIA}\\
D_g^{(n)}:\hspace{0.5cm} & \sum_{j}1 \leq 1-2g \hspace{1.0cm}
&\sum_j(\a_j-1) \leq 3g-1 \;.
\label{homID}
\eea
Here we recognize the homogeneity  requirements~\rf{hom1}
and~\rf{hom2}. The other two requirements are useful for checking the
outcome of the iteration  process.

\newsection{Proof of the conjecture}
To prove that the conjecture is true for $g=1$ we need only to
calculate $W_0(p,p)$ according to~\rf{loopg}. To do this we write the
loop insertion operator as
\beq
\frac{d}{dV(p)}=\frac{\partial}{\partial V(p)}+
                \frac{dc}{dV(p)}\frac{\partial}{\partial c}
\eeq
where
\beq
\frac{\partial}{\partial V(p)}=-\sum_{j=1}^{\infty}\frac{j}{p^{2j+1}}
                             \frac{\partial}{\partial g_j}
\hspace{0.5cm}
\mbox{and}
\hspace{0.5cm}
\frac{dc}{dV(p)}=-\frac{c}{I_1\;(p^2+c)^{3/2}} \;.
\label{dcdV}
\eeq
Noting that
\beq
\frac{\partial}{\partial V(p)}V'(\om) = \frac{2\om p}{(p^2-\om^2)^2}
\eeq
and
\beq
\frac{2p}{(p^2-w^2)^3} = \frac{1}{4p}
\left\{\frac{\partial^2}{\partial p^2} \left(\frac{1}{p^2-\om^2}\right)
-\frac{1}{p}\frac{\partial}{\partial p}
\left(\frac{1}{p^2-\om^2}\right) \right\}
\eeq
one gets
\beq
W_0(p,p)=\frac{c^2}{16p^2(p^2+c)^2}
\eeq
The genus one contribution to the generating functional is thus of the
conjectured form with
\beq
A_1^{(1)}=-\frac{1}{16},\hspace{0.5cm} A_1^{(2)}
= -\frac{c}{16}, \hspace{0.5cm}D_1^{(1)}=\frac{1}{16} \;.
\label{W1coeff}
\eeq

We see these coefficients do not depend on any moments. This is
consistent with the well known property of the 2-loop correlator that its
scaling behaviour is universal~\cite{jm}, {\it i.e.}\ the same for all
multicritical models. Let us mention here that the factor $\frac{1}{p^2}$ in
$W_0(p,p)$ is actually the origin of all divergencies in the continuum
Virasoro generators of the theory. The divergencies appear because one
expands $W_0(p,p)$ in $\frac{1}{a\pi}$ before $a$ is sent to
zero~\cite{metc}. The 2--loop correlator $W_0(p,p)$ itself is however
well defined in the double scaling limit.

To prove the conjecture in the general case we assume that it is true
up to genus $g=g_0$ and calculate the right hand side of the loop
equation~\rf{loopg} for $W_{g_0+1}(p)$. What we would like to show is of
course that
\beq
\hbox{the r.h.s.\ }=\sum_{n=1}^{3(g_0+1)-1} {\cal A}^{(n)} \frac{1}{(p^2+c)^n}
    +\sum_{m=1}^{g_0+1}{\cal D}^{(m)}\frac{1}{p^{2m}}
\label{rhs}
\eeq
where the coefficients ${\cal A}^{(n)}$ and ${\cal D}^{(m)}$
fulfill the homogeneity requirements corresponding to $g=g_0+1$.
It is obvious how one treats the products
$\sum_{g'=1}^{g_0}W_{g'}(p)W_{g_0+1-g'}(p)$. As regards the term
$W_{g_0}(p,p)=\frac{d}{dV(p)}W_{g_0}(p)$ we make use of the fact that the
loop insertion operator when acting on $W_{g_0}(p)$ can be written as
\beq
\frac{d}{dV(p)} = \sum_{i}\frac{dI_i}{dV(p)}\frac{\partial}{\partial I_i}
+\sum_{j}\frac{dM_j}{dV(p)}\frac{\partial}{\partial M_j}
+\frac{dc}{dV(p)} \frac{\partial}{\partial c}
\label{loopio}
\eeq
and that
\bea
\frac{dI_i}{dV(p)}&=&-i\frac{1}{(p^2+c)^{i+1/2}}
                   +(i+1/2)\frac{c}{(p^2+c)^{i+3/2}}
                   -(i+1/2)I_{i+1}\frac{dc}{dV(p)}
\label{dIdV}
\\
\frac{dM_j}{dV(p)}&=&\frac{1}{2}\left\{
-(2j+1)\frac{1}{p^{2j+2}(p^2+c)^{1/2}}-\frac{1}{p^{2j}(p^2+c)^{3/2}}\right\}
\\
&&-\frac{1}{2}\left\{\frac{1}{c}\sum_{l=0}^{j}M_{j-l}\frac{(-1)^l}{c^l}
   +\frac{(-1)^{j+1}}{c^{j+1}}I_1\right\}
   \frac{dc}{dV(p)}
\label{dMdV}
\nonumber
\eea
where $\frac{dc}{dV(p)}$ is given by~\rf{dcdV}.
The details of the general step can be found in Appendix~A.

\newsection{Results for $g=2$ and $g=3$}
It is straightforward to iterate the loop equation~\rf{loopg} starting
from $W_1(p)$ given by~\rf{W1coeff} (and~\rf{conjecture}). To find
$W_g(p)$ one calculates the right hand side of~\rf{loopg} using the
already known lower genus contributions $W_1(p),\ldots,W_{g-1}(p)$ and
the differentiation rules~\rf{dIdV}, \rf{dMdV} and~\rf{dcdV}. This
gives an expression involving fractions of the type
$p^{-2m}(p^2+c)^{-n}$. These fractions can be decomposed into fractions
of the type $p^{-2i}$, $(p^2+c)^{-j}$ allowing one to identify the
coefficients $A_g^{(n)}$ and $D_g^{(m)}$ (cf.\ equations~\rf{Kchi}
and~\rf{KPsi}). The results
for $W_2(p)$ and $W_3(p)$ which are obtained using {\it Mathematica} read

\bea
D_{2}^{(1)}&=&
{{-9}\over {256\,{c^2}\,{{M_{0}}^2}}} - {1\over {64\,{c^2}\,I_{1}\,M_{0}}} +
          {{I_{2}}\over {128\,c\,{{I_{1}}^2}\,M_{0}}}
\nonumber \\
D_{2}^{(2)}&=&
{9\over {256\,c\,{{M_{0}}^2}}}\nonumber\\
A_{2}^{(1)}&=& -D_{2}^{(1)}\nonumber\\
A_{2}^{(2)}&=&-(D_2^{(1)}\,c+D_2^{(2)})\nonumber\\
A_{2}^{(3)}&=&
{{-15}\over {256\,{{I_{1}}^2}}} +
    {{49\,{c^2}\,{{I_{2}}^2}}\over {256\,{{I_{1}}^4}}} +
    {{11\,c\,I_{2} - 20\,{c^2}\,I_{3}}\over {128\,{{I_{1}}^3}}} +
    {5\over {128\,I_{1}\,M_{0}}}\nonumber\\
A_{2}^{(4)}&=&
{{-7\,c}\over {128\,{{I_{1}}^2}}} -
   {{49\,{c^2}\,I_{2}}\over {128\,{{I_{1}}^3}}} \nonumber\\
A_{2}^{(5)}&=&
{{105\,{c^2}}\over {256\,{{I_{1}}^2}}}  \nonumber\\
D_{3}^{(1)}&=&
{{63\,{{I_{2}}^4}}\over {512\,{{I_{1}}^7}\,M_{0}}} -
   {{3\,\left( 9\,{{I_{2}}^3} + 25\,c\,{{I_{2}}^2}\,I_{3} \right) }\over
     {256\,c\,{{I_{1}}^6}\,M_{0}}} \nonumber\\
&& +
   {{201\,{{I_{2}}^2} + 298\,c\,I_{2}\,I_{3} + 145\,{c^2}\,{{I_{3}}^2} +
       308\,{c^2}\,I_{2}\,I_{4}}\over
{2048\,{c^2}\,{{I_{1}}^5}\,M_{0}}}\nonumber\\
&&  +
   {{-85\,c\,I_{2} - 45\,{c^2}\,I_{3} + 38\,M_{0}}\over
     {1024\,{c^4}\,{{I_{1}}^3}\,{{M_{0}}^2}}}\nonumber\\
   & &+
   {{110\,c\,{{I_{2}}^2} - 152\,I_{2}\,M_{0} - 116\,c\,I_{3}\,M_{0} -
       91\,{c^2}\,I_{4}\,M_{0} - 105\,{c^3}\,I_{5}\,M_{0}}\over
     {2048\,{c^3}\,{{I_{1}}^4}\,{{M_{0}}^2}}} \nonumber\\
&& +
   {{153\,\left( 2\,M_{0} + c\,M_{1} \right) }\over
     {2048\,{c^4}\,{{M_{0}}^5}}} +
   {{9\,\left( 14\,M_{0} + c\,M_{1} \right) }\over
     {1024\,{c^4}\,I_{1}\,{{M_{0}}^4}}} \nonumber\\
   & &+
   {{-45\,c\,I_{2}\,M_{0} + 170\,{{M_{0}}^2} - 9\,{c^2}\,I_{2}\,M_{1}}\over
     {2048\,{c^4}\,{{I_{1}}^2}\,{{M_{0}}^4}}}\nonumber\\
D_{3}^{(2)}&=&
{{-27}\over {1024\,{c^3}\,I_{1}\,{{M_{0}}^3}}} +
    {{27\,I_{2}}\over {2048\,{c^2}\,{{I_{1}}^2}\,{{M_{0}}^3}}} -
    {{153\,\left( 2\,M_{0} + c\,M_{1} \right) }\over
      {2048\,{c^3}\,{{M_{0}}^5}}}\nonumber\\
D_{3}^{(3)}&=&
{{225}\over {2048\,{c^2}\,{{M_{0}}^4}}}\nonumber\\
A_{3}^{(1)}&=& -D_3^{(1)}\nonumber\\
A_{3}^{(2)}&=&-(D_3^{(1)}\,c+D_3^{(2)})\nonumber\\
A_{3}^{(3)}&=&
{{5355\,{c^3}\,{{I_{2}}^5}}\over {512\,{{I_{1}}^9}}} +
   {{15\,\left( 171\,{c^2}\,{{I_{2}}^4} - 533\,{c^3}\,{{I_{2}}^3}\,I_{3}
\right) }\over {256\,{{I_{1}}^8}}} \nonumber\\
&&+
   {{1065\,c\,{{I_{2}}^3} - 43010\,{c^2}\,{{I_{2}}^2}\,I_{3} +
       32845\,{c^3}\,I_{2}\,{{I_{3}}^2} + 35588\,{c^3}\,{{I_{2}}^2}\,I_{4}}
      \over {2048\,{{I_{1}}^7}}} \nonumber\\
&&-
   {{155}\over {2048\,{c^2}\,{{I_{1}}^2}\,{{M_{0}}^2}}} +
   {{235\,c\,I_{2} - 64\,M_{0}}\over
     {2048\,{c^2}\,{{I_{1}}^3}\,{{M_{0}}^2}}} +
   {{138\,I_{2} - 19\,c\,I_{3} + 385\,{c^2}\,I_{4}}\over
     {2048\,c\,{{I_{1}}^4}\,M_{0}}} \nonumber\\
&&+
   (840\,c\,{{I_{2}}^3} - 585\,{{I_{2}}^2}\,M_{0} -
       1015\,c\,I_{2}\,I_{3}\,M_{0} \nonumber\\
 &&+ 8975\,{c^2}\,{{I_{3}}^2}\,M_{0}
       19026\,{c^2}\,I_{2}\,I_{4}\,M_{0} -
       14280\,{c^3}\,I_{3}\,I_{4}\,M_{0} \nonumber\\
&& -
       16233\,{c^3}\,I_{2}\,I_{5}\,M_{0})
       /2048\,{{I_{1}}^6}\,M_{0} \nonumber\\
 &&+
   (-18\,{{I_{2}}^2} - 1198\,c\,I_{2}\,I_{3} + 465\,I_{3}\,M_{0} +
       126\,c\,I_{4}\,M_{0} \nonumber\\
&&  - 5439\,{c^2}\,I_{5}\,M_{0} +
       4620\,{c^3}\,I_{6}\,M_{0})/{2048\,{{I_{1}}^5}\,M_{0}} \nonumber\\
&&-
   {{45\,\left( 2\,M_{0} + c\,M_{1} \right) }\over
     {2048\,{c^2}\,I_{1}\,{{M_{0}}^4}}}\nonumber\\
A_{3}^{(4)}&=&
{{-5355\,{c^3}\,{{I_{2}}^4}}\over {256\,{{I_{1}}^8}}} +
    {{3\,\left( -3105\,{c^2}\,{{I_{2}}^3} +
          7279\,{c^3}\,{{I_{2}}^2}\,I_{3} \right) }\over {512\,{{I_{1}}^7}}}
\nonumber\\
 &&    + {{-1590\,c\,{{I_{2}}^2} + 44116\,{c^2}\,I_{2}\,I_{3} -
        17545\,{c^3}\,{{I_{3}}^2} - 37373\,{c^3}\,I_{2}\,I_{4}}\over
      {2048\,{{I_{1}}^6}}} \nonumber\\
&&  - {{315}\over
      {2048\,c\,{{I_{1}}^2}\,{{M_{0}}^2}}} -
    {{21}\over {512\,c\,{{I_{1}}^3}\,M_{0}}} +
    {{7\,\left( -21\,I_{2} + 125\,c\,I_{3} \right) }\over
      {2048\,{{I_{1}}^4}\,M_{0}}} \nonumber\\
&& +
    (-1428\,c\,{I_{2}}^2 + 945\,I_{2}\,M_{0} + 131\,c\,I_{3}\,M_{0} -
        11109\,{c^2}\,I_{4}\,M_{0} \nonumber\\
 &&+ 10185\,{c^3}\,I_{5}\,M_{0})/2048\,{I_{1}}^5\,M_{0}\nonumber\\
A_{3}^{(5)}&=&
{{7371\,{c^3}\,{{I_{2}}^3}}\over {256\,{{I_{1}}^7}}} +
    {{47196\,{c^2}\,{{I_{2}}^2} - 69373\,{c^3}\,I_{2}\,I_{3}}\over
      {2048\,{{I_{1}}^6}}} \nonumber\\
  &&+ {{1566\,c\,I_{2} - 22029\,{c^2}\,I_{3} +
        17465\,{c^3}\,I_{4}}\over {2048\,{{I_{1}}^5}}} \nonumber\\
&& +
    {{231}\over {2048\,{{I_{1}}^3}\,M_{0}}} +
    {{21\,\left( 77\,c\,I_{2} - 45\,M_{0} \right) }\over
      {2048\,{{I_{1}}^4}\,M_{0}}}\nonumber\\
A_{3}^{(6)}&=&
{{-99\,c}\over {2048\,{{I_{1}}^4}}} -
   {{64295\,{c^3}\,{{I_{2}}^2}}\over {2048\,{{I_{1}}^6}}} +
   {{11\,\left( -3813\,{c^2}\,I_{2} + 2755\,{c^3}\,I_{3} \right) }\over
     {2048\,{{I_{1}}^5}}}\nonumber\\
&& - {{1155\,c}\over {2048\,{{I_{1}}^3}\,M_{0}}}\nonumber\\
A_{3}^{(7)}&=&
{{21021\,{c^2}}\over {2048\,{{I_{1}}^4}}} +
    {{13013\,{c^3}\,I_{2}}\over {512\,{{I_{1}}^5}}}\nonumber\\
A_{3}^{(8)}&=&
{{-25025\,{c^3}}\over {2048\,{{I_{1}}^4}}}
\nonumber
\eea
We note that $A_g^{(1)}=-D_g^{(1)}$ and $A_g^{(2)}=-(D_g^{(1)} \,c
+D_g^{(2)})$ for both $g=2$ and $g=3$. The same is true for
$g=1$ according to~\rf{W1coeff}. It is easy to show
that the relations hold
for any genus.
Going through the induction proof in Appendix~A one finds that
$p^{-2}$, $p^{-4}$, $(p^2+c)^{-1}$ and $(p^2+c)^{-2}$ never appear in
isolation on the right hand side of the loop equation~\rf{loopg} but
only in the combinations $p^{-2}(p^2+c)^{-1}$, $p^{-2}(p^2+c)^{-2}$,
$p^{-4}(p^2+c)^{-1}$ and $p^{-4}(p^2+c)^{-2}$. Decomposing these
fractions one gets the relations between the coefficients stated above.

The expressions for the $A$'s and $D$'s look rather
complicated. However, many terms in $W_g(p)$
can be ignored in the double scaling
limit. In order for a given term to survive the double scaling limit
the equality sign must hold in both homogeneity relations in~\rf{homIA}
for $A$ terms and in~\rf{homID} for $D$ terms. From the list
of coefficients given above and from~\ref{W1coeff} one immediately finds
that for $g=1$, $g=2$ and $g=3$
all $D$ terms disappear. It actually holds that
the second sum in~\rf{conjecture} vanishes in the double scaling limit
for all genera.
This
follows from the fact that~\rf{hom1} is always fulfilled and that
$p^{-2m}$ never appears in isolation on the right side of~\rf{loopg}
but always in combination with at least one power of $(p^2+c)^{-1}$.
In fact the induction proof can be carried trough with the homogeneity
requirement $H_I(d_g^{(m)})\leq 3g-2$ in stead of $H_I(d_g^{(m)})\leq
3g-1$.

\newsection{Discussion}
We have shown that it is possible to calculate for the complex matrix
model any multi--loop correlator to any  genus. This gives us the
possibility of going beyond the planar approximation which has been used
in most calculations to date. For example it is interesting to note
that in the double scaling limit
$W_1(p)$ reduces to
\beq
W_1= -\frac{c}{16}\chi^{(2)} \hspace{1.0cm} (d.\,s.\,l.)
\eeq
which apart from a trivial factor is exactly the same as for the reduced
hermitian matrix model \cite{jjm}:
\beq
W_1^{C}(\pi,\Lambda)=\frac{1}{4}W_1^{H}(\pi,\Lambda)\;.
\eeq
It was conjectured long time ago that
$W_g^{C}(\pi,\Lambda)=\frac{1}{4^g}W_g^{H}(\pi,\Lambda)$ \cite{jjm}.
The iterative procedure described above and the similar
procedure in the hermitian matrix model case provide us with the
possibility of testing this conjecture in detail. Presumably it is even
possible within the iterative framework to give a rigorous proof of the
conjecture. The iterative scheme also allows us to study higher genus
contributions to the multi--loop correlators, for instance to
investigate whether these can be expressed in a closed form as was the
case for genus zero.

\setcounter{section}{0}

\appendix{Details of the proof}
We now go through the general step of the induction proof following the
line of action described in section~5. Hence let us assume that the
conjecture is true up to genus $g=g_0$ and let us calculate the right
hand side of the loop equation~\rf{loopg} for $W_{g_0+1}(p)$.
We consider first the products
$\sum_{g'=1}^{g_0}W_{g'}(p)W_{g_0+1-g'}(p)$
For this purpose it is convenient to rewrite the basis vectors
$\Psi^{(n)}(p)$ and $\chi^{(n)}(p)$ as
\bea
\Psi^{(n)}(p)&=&\sum_{i=1}^{n}\Omega^{(i)}(p)\;
P_i^{(n)}(M_0,\ldots,M_{n-i})\\
\chi^{(n)}(p)&=& \sum_{i=1}^{n}\Phi^{(i)}(p)\;
Q_i^{(n)}(I_1,\ldots,I_{n-i+1})
\eea
where
\beq
H_M(P_i^{(n)})=n-i, \hspace{1.0cm} H_I(Q_i^{(n)})=n-i
\eeq
and
\beq
H_M(Q_i^{(n)})=H_I(P_i^{(n)})=0 \;.
\eeq

The sum above consists of three types of terms: the one which products of
$\chi$'s enter, the one which products of $\Psi$'s enter and the one which
contains the mixed products $\chi \cdot \Psi$. The first of these gives
rise to only terms of the type $K^{(n)}\,(p^2+c)^{-n}$, $n\in [3,3g_0+1]$
thus contributing only to the first sum in~\rf{rhs}. The constant
$K^{(n)}$ is a sum (over $g'$, $q$, $p$ and $i$) of products with the
following structure
\[
K^{(n)}\sim  A_{g'}^{(q)}\;A_{g_0+1-g'}^{(p)}\;Q_i^{(q)}\;Q_{n-i-1}^{(p)}
\]
It is easily seen that
\[
H_I(K^{(n)})\leq 3g_0+2-n=3(g_0+1)-n-1,\hspace{1.0cm}
H_M(K^{(n)})\leq g_0+1 \;.
\]
The second type of terms, those which contain products of $\Psi$'s give
rise to fractions of the type $C^{(m)}\,p^{-2m}(p^2+c)^{-1}$, $m\in
[2,g_0+1]$. Decomposing a fraction like this one gets a sum of fractions
$p^{-2k}$, $k \in [1,m]$ plus the fraction $(p^2+c)^{-1}$, all of them
with some weight which depends only on c. The constant $C^{(m)}$ is a
sum (over $g'$, $q$, $p$ and $i$) of products with the following
structure.
\[
C^{(m)}\sim D^{(q)}_{g'}\;D_{g_0+1-g'}^{(p)}\;P_i^{(q)}\; P_{m-i}^{(p)}\;.
\]
One finds
\[
H_M(C^{(m)})\leq g_0+1-m\hspace{0.5cm}\mbox{and}\hspace{0.5cm}
H_I(C^{(m)}) \leq 3(g_0+1)-2
\]
which means that the homogeneity requirements~\rf{hom1} and~\rf{hom2}
are fulfilled.
Finally, let us consider the term with the mixed products
$\chi\cdot\Psi$. From this one we get fractions of the type
$B^{(n,m)}p^{-2m}(p^2+c)^{-n}$, $m\in [1,g_0]$, $n \in [2,3g_0]$. After
decomposition we are left with a sum of fractions of the type $p^{-2i}$,
 $i \in [1,m]$ and $(p^2+c)^{-j}$, $j\in [1,n]$. Here the constant
$B^{(n,m)}$ is a sum (over $g'$, $q$ and $p$) of products like the
following.
\[
B^{(n,m)}\sim D_{g'}^{(q)}\; A_{g_0+1-g'}^{(p)}\;P_m^{(q)}\;Q_{n-1}^{(p)}
\]
and one can show that
\[
H_I(B^{(n,m)})\leq 3(g_0+1)-1-n,\hspace{1.0cm} H_M(B^{(n,m)})\leq g_0+1-m
\]
which means that~\rf{hom1} and~\rf{hom2} are satisfied also in this
case. This completes the treatment of the first term on the right hand
side of the loop equation. Let us now turn to the term
$W_{g_0}(p,p)=\frac{d}{dV(p)} W_{g_0}(p)$. We start by recalling
that the loop insertion operator when acting on $W_{g_0}(p)$ can be
written as in~\rf{loopio}--\rf{dMdV}.
Furthermore it is convenient to write $W_{g_0}(p)$ in the following
form.
\beq
W_{g_0}(p)=\sum_{i=1}^{3g_0-1}\hat{A}_{g_0}^{(i)} \Phi^{(i)}(p)
          +\sum_{j=1}^{g_0}\hat{D}_{g_0}^{(j)}\Omega^{(j)}(p)
\label{loopalt}
\eeq
where
\bea
\hat{A}^{(i)}_{g_0}&=&\hat{A}^{(i)}_{g_0}(M_0,\ldots
M_{g_0},I_1,\ldots, I_{3g_0-i})=\sum_{n=i}^{3g_0-1}A_{g_0}^{(n)}
Q^{(n)}_i\\
\hat{D}^{(j)}_{g_0}&=&\hat{D}^{(j)}_{g_0}(M_0,\ldots
M_{g_0-j},I_1,\ldots, I_{3g_0})=\sum_{m=j}^{g_0}D_{g_0}^{(m)}
P^{(m)}_j
\eea
and
\bea
H_I(\hat{A}_{g_0}^{(i)})&\leq& 3g_0-i-1,\hspace{1.0cm}
H_M(\hat{A}_{g_0}^{(i)})\leq g_0 \\
H_I(\hat{D}_{g_0}^{(j)})&\leq &3g_0-1,\hspace{1.3cm}
H_M(\hat{D}_{g_0}^{(j)})\leq g_0-j
\eea

Differentiation of $\Phi^{(i)}(p)$ and $\Omega^{(j)}(p)$ is simple since
these expressions depend only on c. One finds
\bea
\frac{d\Phi^{(j)}(p)}{dV(p)} &= &
(j+\frac{1}{2})\,\frac{c}{I_1}\;\frac{1}{(p^2+c)^{j+3}} \label{phid} \\
\frac{d\Omega^{(j)}(p)}{dV(p)} &=&
\frac{c}{2I_1}\;\frac{1}{p^{2j}(p^2+c)^3} \;.
\label{omd}
\eea
Differentiation of the $\Phi$'s in~\rf{loopalt} thus gives rise to
terms of the type $K^{(n)} (p^2+c)^{-n}$, $n\in [4,3(g_0+1)-1]$ where
\[
K^{(n)}\propto \hat{A}_{g_0}^{(n-3)}
\]
while differentiation of the $\Omega$'s in~\rf{loopalt} gives terms of
the type $C^{(m)}\,(p^2+c)^{-3}p^{-2m}$, $m \in [1,g_0]$ where
\[
C^{(m)}\propto \hat{D}_{g_0}^{(m)} \;.
\]
It is easy to see that~\rf{hom1} and~\rf{hom2} are satisfied in both
cases.
The partial differentiation of $\hat{A}^{(i)}_{g_0}$ and
$\hat{D}_{g_0}^{(j)}$ with respect to $c$ results in terms which can be
obtained from those just mentioned (except for trivial factors) by
decreasing the power of $(p^2+c)^{-1}$ by one and replacing the coefficients
by their derivatives with respect to $c$. Since differentiation with
respect to $c$ does not change the degree of homogeneity, it follows
that all resulting terms are in agreement with the conjecture.

We now turn to the partial differentiation of the $\hat{D}$'s and
$\hat{A}$'s with respect to the moments $I_j$. It is obvious that we do
not have to worry about terms arising from the first term in~\rf{dIdV}
as long as we check that everything which comes from the second one is
ok. However, checking the second one also takes care of the third. This
is so because each term originating from the third term in~\rf{dIdV}
has a brother term originating from the second one from which it
differs only by $I_{i+1}(p^2+c)^{-i}$ (apart from trivial factors).

Taking the partial derivative of the $\hat{A}$'s with respect to the
$I$'s we get from the second term in~\rf{dIdV} terms
of the type $K^{(n)}\;(p^2+c)^{-n}$, $n\in[4,3(g_0+1)-1]$ where
\[
K^{(n)}=\sum_{i=1}^{3g_0-1}\frac{\partial \hat{A}_{g_0}^{n-i-2}}
        {\partial I_i} (i+\frac{1}{2})
\]
for which it holds that
\[
H_I(K^{(n)})\leq 3(g_0+1)-n-1, \hspace{1.0cm} H_M(K^{(n)})\leq g_0
\]
where we have used that differentiating $\hat{A}_{g_0}^{(i)}$ with
respect to $I_i$ lowers $H_I$ by $i-1$.

The terms which originate from the second term in~\rf{dIdV} when we
take the partial derivative of the $\hat{D}$'s with respect to the $I$'s
are of the type $B^{(n,m)}\;p^{-2m}(p^2+c)^{-n}$, $m\in [1,g_0]$,
$n\in[3,3(g_0+1)-1]$. The constants are given by
\[
B^{(n,m)}\propto \frac{\partial \hat{D}_g^{(m)}}{\partial I_{n-2}} \;.
\]
It is easy to show that
\[
H_I(B^{(n,m)}) \leq 3(g_0+1)-m-1,\hspace{1.0cm} H_M(B^{(n,m)})\leq g_0-n
\]
which means that~\rf{hom1} and~\rf{hom2} are satisfied for all terms.
Now we only have left the partial differentiation of the $\hat{A}$ and
the $\hat{D}$'s with respect to the $M$'s. In this case checking terms
arising from the first line in~\rf{dMdV} automatically takes care of
terms arising from the second one. This is obvious for the last term in
the second line.
For the sum it follows from the fact that its most problematic term,
the one with the $M_j$, has a brother term in the first line where $M_j$
is replaced by $p^{2j}$ (modulo some trivial factors).

Taking the partial derivative of the $\hat{A}$'s with respect to the
$M$'s we get from the first term in~\rf{dMdV} terms of the type
$B^{(n,m)}\;p^{-2m}(p^2+c)^{-n}$, $m\in [1,g_0+1]$,
$n\in[2,3g_0]$ where
\beq
B^{(n,m)}\propto\frac{\partial \hat{A}_{g_0}^{(n-1)}}{\partial
M_{m-1}} \;.
\label{Bnm}
\eeq
For $B^{(n,m)}$ it holds that
\beq
H_I(B^{(n,m)})\leq 3(g_0+1)-n-3,\hspace{1.0cm} H_M(B^{(n,m)})\leq g_0+1-m
\label{HBnm}
\eeq
where we have used that differentiating once with respect to $M_j$
lowers $H_M$ with $j$. Hence~\rf{hom1} and~\rf{hom2} are satisfied.
The second term in~\rf{dMdV} gives rise to terms of the type
$B^{(n,m)}\;p^{-2(m-1)}(p^2+c)^{-n-1}$, $m\in [1,g_0+1]$,
$n\in[2,3g_0]$ where $B^{(n,m)}$ is given by~\rf{Bnm}. It follows
from~\rf{HBnm} that the homogeneity conditions are satisfied also in
this case.

Differentiating the $\hat{D}$'s partially with respect to the $M$'s we
get from the first term in~\rf{dMdV} terms of the type
$K^{(m)}\;p^{-2m}(p^2+c)^{-1}$, $m\in [3,g_0+1]$ where
\beq
K^{(m)}=-\sum_{i=0}^{g_0}\frac{\partial
\hat{D}_{g_0}^{(m-i-1)}}{\partial M_i} (i+\frac{1}{2}) \;.
\label{Kn}
\eeq
One finds
\beq
H_I(K^{(m)}) \leq 3g_0-1 \hspace{1.0cm} H_{M} \leq g_0+1-m
\label{HKn}
\eeq
which is in accordance with~\rf{hom1} and~\rf{hom2}. From the second
term in~\rf{dMdV} we get terms of the type
$K^{(m)}\;p^{-2(m-1)}(p^2+c)^{-2}$, $m\in [3,g_0+1]$ where
$K^{(n)}$ is given by~\rf{Kn}. It follows from~\rf{HKn} that the
homogeneity conditions are fulfilled also in this case. This completes
the proof of our conjecture.

\appendix{Relation to external field problem}
Let us now show that the complex matrix model
considered so far is equivalent to
an external field problem for the complex matrix model with a
logarithmic potential.
We start from the partition function
\beq
Z[\eta^{\dagger}\eta]= e^{-N\tr(\eta^{\dagger}\eta)}
\int d\phi^{\dagger} d\phi \exp \left\{N \tr
(- \phi^{\dagger} \phi+\phi^{\dagger}\eta +\eta^{\dagger}\phi
+\alpha \log(\phi^{\dagger} \phi))\right\}
\label{Z-eta}
\eeq
where $\phi$ and $\eta$ are complex $N\times N$ matrices. Exploiting the
invariance of the measure of integration, one finds in the standard way the
following equation of motion
\beq
\left\{-\frac{\partial^{2}}{\partial \eta^{\dagger}_{ji}
\partial \eta_{ik}}+2\alpha N^2\delta_{jk}
-N\eta_{ij}\frac{\partial}{\partial\eta_{ik}}
-N\eta^{\dagger}_{ki}\frac{\partial}{\partial\eta^{\dagger}_{ji}}
\right\} Z[\eta^{\dg}\eta]=0 \;.
\label{sd}
\eeq
Since the partition function (\ref{Z-eta}) depends only on
(positive) eigenvalues of $\eta^{\dagger}\eta$ which we denote as
$\lambda_i^2$,
one can derive from this equation the so-called master equation of the
theory
\beq
\left\{-\frac{1}{2}\frac{1}{\lambda_i}\frac{\partial}
{\partial \lambda_i}
\lambda_i\frac{\partial}{\partial \lambda_i}
- \sum_{j\neq i}\frac{\lambda_i\frac{\partial}{\partial \lambda_i}
-\lambda_j\frac{\partial}{\partial \lambda_j}}
{{\lambda_i}^2-\lambda_j^2}-2N\lambda_i\frac{\partial}{\partial
\lambda_i} +2\alpha N^2 \right\}
Z[\lambda^2]=0 \;.
\label{master}
\eeq

Let us introduce the new variables $t_k$ which are related to  the external
field $\eta$ by a kind of the Miwa transformation
\bea
t_k&=&\frac{1}{k} \tr(\eta^{\dagger}\eta)^{-k} + 2N \delta_{k,1}
\hspace{1.0cm} k>0 \\
t_0&=&\tr \log(\eta^{\dagger}\eta)^{-1} \;.
\eea
Using the chain rule one can write~\rf{sd} (or \rf{master}) as
\beq
\sum_{n=-1}^{\infty}(\eta^{\dagger}\eta)^{-n-1} L_n
Z[\eta^{\dagger}\eta]=0
\eeq
where
\bea
L_{-1}&=& \a N^2-2N\frac{\partial}{\partial t_0} \label{L_1} \\
L_{n}&=& \sum_{k=0}^{n}\frac{\partial}{\partial t_k}
\frac{\partial}{\partial t_{n-k}}
+\sum_{k=0}^{\infty}kt_k\frac{\partial}{\partial t_{n+k}}
\hspace{1.0cm} n\geq 0 \label{L_n} \;.
\eea

In the limit $N\rightarrow \infty$ where the traces become
independent variables we recover from these equations the Virasoro
constraints \cite{jjm} of the matrix model
\beq
Z=\int d\phi^{\dg} d\phi\exp(-\sum_{k=0}^{\infty}t_k\tr(\phi^{\dg} \phi)^k)
\eeq
where the integration is over complex $\frac{\alpha N}{2} \times
\frac{\alpha N}{2}$ matrices. This matrix model is trivially related
to the one defined by~\rf{partition} and~\rf{potential}.

Similarly to \cite{cm1} it can be shown that the model (\ref{Z-eta}) is
equivalent to a complex-matrix analog of the Kontsevich--Penner model
defined by the partition functon
\beq
Z[\Lambda^\dagger \Lambda] =
\int d\phi^{\dagger} d\phi \exp \left\{N \tr
(- \Lambda \phi^{\dagger} \Lambda^\dagger\phi
+\alpha \left[\log(1+\phi^{\dagger})(1+ \phi)
-\phi^{\dagger} -\phi\right])\right\} \;.
\label{Z-Lambda}
\eeq
This partition function can be obtained (modulo a \sloppy
$\Lambda$-dependent factor) from~\rf{Z-eta} by substitution
$\phi\rightarrow (\Lambda^\dagger)^{\frac 12} \phi (\Lambda)^{\frac 12}+
(\Lambda^\dagger \Lambda)^{\frac 12}$ providing
\beq
\eta=(\Lambda^\dagger\Lambda)^{\frac 12}-\alpha
(\Lambda^\dagger\Lambda)^{-\frac 12} \;.
\eeq


\begin{thebibliography}{99}
\bibitem{Morris1}T. R. Morris, Nucl. Phys. B356 (1991) 703
\bibitem{jjm}J. Ambj\o rn, J. Jurkiewicz and Yu. Makeenko, Phys. Lett.
B251 (1990) 517
\bibitem{Morris2} S. Dalley, C. Johnson and T. Morris, {\sl
Multicritical Complex Matrix Models and Nonperturbative 2D Quantum
Gravity}, SHEP 90/91-16, and {\sl Nonperturbative Two--Dimensional
Quantum Gravity}, SHEP 90/91-28
\bibitem{metc}Yu. Makeenko, A. Marshakov, A. Mironov and A. Morozov,
Nucl. Phys. B356 (1991) 574
\bibitem{Migdal}A. A. Migdal, Phys. Rep. 102 (1983) 199
\bibitem{david}F. David, Mod. Phys. Lett. A5 (1990) 1019
\bibitem{jm} J. Ambj\o rn and Yu. Makeenko, Mod. Phys. Lett. A5 (1990)
1753
\bibitem{jcm}J. Ambj\o rn, L. Chekhov and Yu. Makeenko, {\sl Higher
Genus Correlators and W-infinity from the Hermitian One-Matrix-Model},
NBI-HE-92-22
\bibitem{cm}L. Chekhov and Yu. Makeenko, Phys. Lett. B278 (1992) 271
\bibitem{cm1} L. Chekhov and Yu. Makeenko, Mod. Phys. Lett. A7 (1992) 1223
\bibitem{Mak0}Yu. Makeenko, Pis'ma v ZhETF 52 (1990) 885
\end{thebibliography}
\end{document}